\documentclass[aps,pra,onecolumn,groupedaddress,notitlepage,10pt]{revtex4-1}

\usepackage[T1]{fontenc}
\usepackage{graphicx}
\usepackage[labelfont=bf]{caption}
\usepackage{color}
\usepackage{hyperref}
\usepackage{amsmath}
\usepackage{amssymb}
\usepackage{setspace}

\newcommand{\ee}{\mathrm{e}}
\newcommand{\ii}{\mathrm{i}}
\newcommand{\dirac}[1]{\left< #1 \right>}

\begin{document}


\title{Super-resolved multimodal multiphoton microscopy with spatial frequency-modulated imaging}


\author{Jeffrey J.~Field, Keith W.~Wernsing, Scott R.~Domingue}
\affiliation{Department of Electrical and Computer Engineering, Colorado State University, Fort Collins, CO 80523 \\ \href{mailto:jeff.field@colostate.edu}{\color{blue} jeff.field@colostate.edu}}

\author{Alyssa M.~Allende Motz} 
\affiliation{Department of Physics, Colorado School of Mines, Golden, CO 80401}
\affiliation{National Renewable Energy Laboratory, Golden, CO 80401}

\author{Keith F.~DeLuca, Jennifer G.~DeLuca}
\affiliation{Department of Biochemistry and Molecular Biology, Colorado State University, Fort Collins, CO 80523}

\author{Darius Kuciauskas, Dean H.~Levi}
\affiliation{National Renewable Energy Laboratory, Golden, CO 80401}

\author{Jeff A.~Squier}
\affiliation{Department of Physics, Colorado School of Mines, Golden, CO 80401}

\author{Randy A.~Bartels}
\affiliation{Department of Electrical and Computer Engineering, Colorado State University, Fort Collins, CO 80523}
\affiliation{School of Biomedical Engineering, Colorado State University, Fort Collins, CO 80523}

\date{\today}

\begin{abstract}
Super-resolved far-field microscopy has emerged as a powerful tool for investigating the structure of objects with resolution well below the diffraction limit of light \cite{Gustafsson:2005,Betzig:2006,Hell:2007}.  Nearly all super-resolution imaging techniques reported to date rely on real energy states of probe molecules to circumvent the diffraction limit, preventing super-resolved imaging of contrast mechanisms that occur via virtual energy states such as harmonic generation (HG). Here we report a super-resolution technique based on SPatIal Frequency modulated Imaging (SPIFI) \cite{Futia:2011,Schlup:2011,Higley:2012} that permits super-resolved nonlinear microscopy with any contrast mechanism, and with single-pixel detection. We show multimodal super-resolved images with two-photon excited fluorescence (TPEF) and second-harmonic generation (SHG) from biological and inorganic media. Multiphoton SPIFI (MP-SPIFI) \cite{Hoover:2012} provides spatial resolution up to 2$\eta$ below the diffraction limit, where $\eta$ is the highest power of the nonlinear intensity response. MP-SPIFI has the potential to not only provide enhanced resolution in optically thin media, but shows promise for providing super-resolved imaging at depth in scattering media -- opening the possibility of \textit{in vivo} super-resolved imaging.
\end{abstract}


\maketitle

\section{Introduction}
Of the numerous far-field image formation techniques, multiphoton microscopy (MPM) has proved particularly valuable in many biological studies, providing multimodal images with robustness to optical scattering \cite{Helmchen:2005}, and label-free images of complex biological structures \cite{Zoumi:2002,Olivier:2010,Campagnola:2011} and inorganic harmonic probes \cite{Pantazis:2010,Grange:2011}. Scattering robustness is achieved twofold with nonlinear excitation. First, inherent optical sectioning provided by the nonlinear intensity dependence \cite{Denk:1990} allows for collection of signal light with a single-pixel detector. Second, the use of excitation wavelengths in the near infrared (NIR) \cite{Kobat:2009,Horton:2013} enhance penetration depth of the ballistic photons responsible for nonlinear contrast generation \cite{Dunn:2000}. Due to the diffraction limit of light, these longer excitation wavelengths yield lower spatial resolution than wavelengths in the visible spectral region, such as those used for confocal fluorescence microscopy. 

The most successful super-resolution imaging techniques reported to date have exploited the real energy states of probe molecules to generate image contrast.  These real energy states provide a means to manipulate how contrast is generated, thereby reducing the effective size of the point-spread function (PSF) which sets image resolution. Some examples include photoactivated localization microscopy (PALM) \cite{Betzig:2006,Betzig:1995}, stimulated emission depletion (STED) microscopy \cite{Hell:2007,Hell:1994,Hell:2003}, ground state depletion (GSD) microscopy \cite{Hell:1995_02,Hell:2007_02}, saturation of transient absorption in electronic states \cite{Wang:2013}, and saturation of scattering due to surface plasmon resonance \cite{Chu:2014:01,Chu:2014:02}.

Conversely, the instantaneous response of the contrast medium in coherent nonlinear scattering results in no net energy transfer. In this case, the photokinetics are often described via virtual energy states. The application of most super-resolution imaging methods to virtual energy states is not possible. Despite this, examples of HG microscopy with sub-diffraction-limited resolution have been reported \cite{Masihzadeh:2009,Liu:2014}. While these methods succeeded in collecting HG images with sub-diffraction limited resolution, both are unlikely to be robust to imaging at depth in tissue due to effects such as birefringence and circular dichroism.

Until now, a technique that provides super-resolved imaging of any contrast mechanism in complex media has not yet emerged. Here we report the use of MP-SPIFI for super-resolved imaging of SHG and TPEF. Combining the scattering robustness offered by illumination wavelengths in the NIR with single-pixel detection of signal light, MP-SPIFI has the potential to provide super-resolved images at depth in scattering media. 

\section{Physical basis of resolution enhancement}
The underlying principle of MP-SPIFI is to illuminate the specimen with a spatio-temporally varying intensity pattern, $I_\mathrm{ill}({\bf r},t)$, that contains a discrete set of well-separated spatial frequency components. The fundamental lateral ($x$) spatial frequency of the illumination intensity in the object region, $f_{x,2}$, is varied by modulating an incident laser beam with an amplitude grating whose fundamental spatial frequency, $f_{x,1}$, varies with time. Signal light, $\beta({\bf r},t)$, generated by the interaction of probe molecules in the specimen with the illuminating intensity, is collected on a single-element photodetector such as a photomultiplier tube (PMT). The signal from the photodetector is expressed mathematically as:
\begin{equation}
S(t) = \left< \beta({\bf r},t) \right>_{\bf r} =  \left< \left[I_\mathrm{ill}({\bf r},t)\right]^\eta \, C({\bf r},t) \right>_{\bf r}
\label{eq:signal}
\end{equation} 
where $C({\bf r},t)$ is the spatiotemporal distribution of probe molecules, including parameters such as cross-section, quantum efficiency, etc., and $\left< \, \cdot \,  \right>_{\bf r}$ denotes integration over all space. 

Physically, Eq.~\eqref{eq:signal} represents a projection of the spatial frequencies present in the illumination intensity onto the object, thereby encoding the amplitude of these spatial frequency components that are present in $C({\bf r},t)$. Since $f_{x,2}$ is varied through the entire span of frequency support provided by the illumination system, $S(t)$ encodes a one-dimensional (1D) representation of the specimen in the lateral spatial frequency domain.

\begin{figure*}
\begin{center}
\resizebox{\linewidth}{!}{\includegraphics{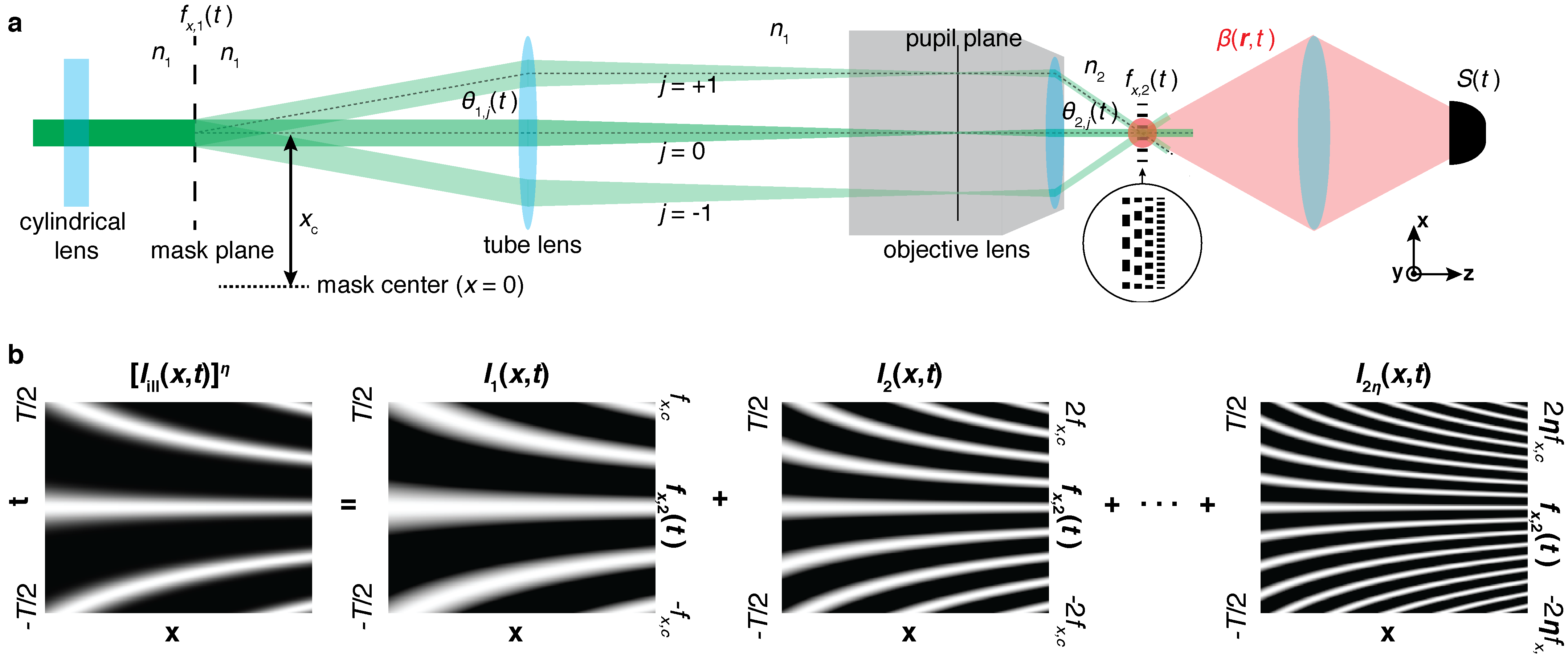}}
\caption{\label{fig:schematic} Physical principles of super-resolved imaging with MP-SPIFI. (a) Schematic representation of the imaging system. (b) A decomposition of the nonlinear illumination intensity into constituent harmonics of spatial frequency. DC spatial frequencies have been omitted for simplicity. $T = 1/\nu_r$ is the period of mask rotation.}
\end{center}
\end{figure*}

A schematic of the MP-SPIFI microscope in the lateral-axial plane is shown in Fig.~\ref{fig:schematic}a. Schematics showing the microscope in both the lateral-axial and the vertical-axial planes, $(x,z)$ and $(y,z)$ respectively, are shown in Fig.~S\ref{sfig1}. In the mask plane, a spinning disk with rotational frequency $\nu_r$ imparts a time-dependent lateral spatial frequency, $f_{x,1}(t)$, that varies linearly with time, causing the vertically-focused pulse train incident on the mask to diffract into several beams with varying propagation angles, $\theta_{1,j}(t)$, where $j$ is the diffracted order.

In the object plane, conjugate to the mask plane, the diffracted orders interfere to form a spatial intensity pattern containing a discrete set of spatial frequencies that vary linearly with scan time. The spatial frequencies contained in the illumination intensity at any time $t$ are harmonics of the fundamental spatial frequency imparted by the mask. In the case drawn here, where $j\in \{-1,0,1\}$, the spatial frequencies in the intensity pattern are $\{0, f_{x,2}(t), 2 f_{x,2}(t)\}$, where $f_{x,2}(t) = (n_2/n_1) \, M \, f_{x,1}(t)$, $n_1$ and $n_2$ are the indices of refraction in the mask and object planes respectively, and $M$ is the magnification of the imaging system, defined as the ratio of the geometric focal lengths of the tube lens and the objective lens, $M = F_t/F_o$.

The fundamental spatial frequency in the object region, $f_{x,2}(t)$, appears in the intensity due to interference of the $j = \{0,1\}$ and $j = \{0,-1\}$ beams, while the second-harmonic of the spatial frequency, $2 \, f_{x,2}(t)$, appears due to the interference of the $j = \pm1$ beams. In a linear-excitation SPIFI microscope, $S(t)$ encodes {\em two} images of the specimen \cite{Futia:2011}: one with lateral spatial frequency support limited by the cutoff frequency of the objective lens, defined as $f_{x,c} = \mathrm{NA}/\lambda$, and a second image with spatial frequency support of $2 f_{x,c}$. 

When the signal light $\beta({\bf r},t)$ depends nonlinearly on the illumination intensity, the frequency support range is expanded further by additional harmonics of $f_{x,2}(t)$ that appear due to the spatial nonlinearity. For example, nonlinearities in fluorescence intensity arising from saturation of linear electronic absorption, similar to saturated structured illumination microscopy (SSIM) \cite{Gustafsson:2005}, could be used to project spatial frequencies beyond those supported by the objective lens. In this case the signal light can be represented as $\beta({\bf r},t) = f\left[I_\mathrm{ill}({\bf r},t) \right] C({\bf r},t)$, and the functional form $f[\cdot]$ can be expanded in a Taylor series to find the relative weights of higher-order spatial frequency components \cite{Heintzmann:2009}. 

Similarly, nonlinearities arising from multiphoton contrast lead to additional spatial frequency support. Here we consider nonlinearities of integer-order $\eta$, such that $\beta({\bf r},t) = \left[I_\mathrm{ill}({\bf r},t) \right]^\eta C({\bf r},t)$. Consequently, the temporal signal collected from the PMT encodes a set of 2$\eta$ images with spatial frequency support up to 2$\eta$-times the cutoff of the objective lens. This can be easily shown by examining the spatio-temporal illumination intensity in the lateral dimension ($x$) with scan time (Supplementary Information): 
\begin{equation}
\left[ I_\mathrm{ill}(x,t) \right]^\eta = \sum \limits_{q = 0}^{2 \eta} b_q(t) \, \cos \left[ 2 \pi \, q \, f_{x,2}(t) \, x + q \, \nu_c \, t \right] = \sum \limits_{q = 0}^{2 \eta} I_q(x,t)
\end{equation}
where $\nu_c \equiv (n_2/n_1) \, M \, \Delta k \, \nu_r \, x_c$ is a carrier frequency that arises from the non-zero modulation frequency provided by the mask on the optic axis of the microscope (Fig.~\ref{fig:schematic}a), $\Delta k$ sets the density of the modulation mask \cite{Futia:2011}. 

\begin{figure}
\begin{center}
\resizebox{0.5\linewidth}{!}{\includegraphics{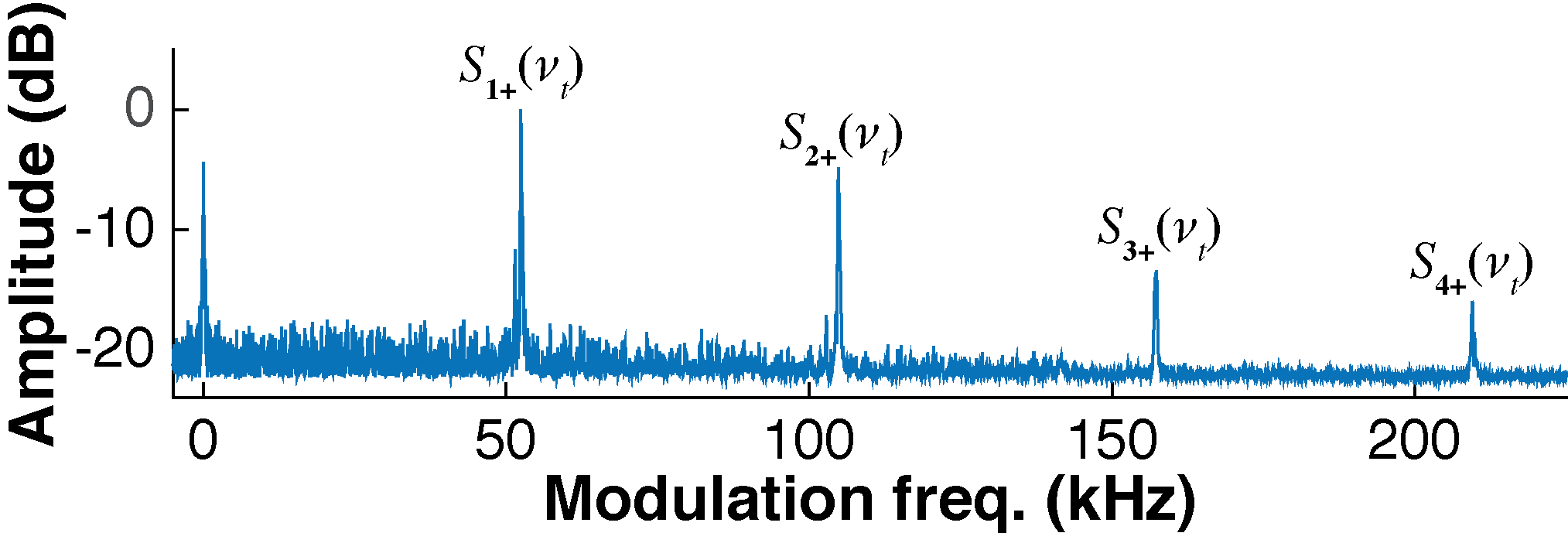}}
\caption{\label{fig:signals}  SHG-SPIFI signal measured from a single BaTiO$_3$ particle. Each SHG-SPIFI image, $S_{q+}(\nu_t)$, is centered about a harmonic of the carrier frequency, allowing for separation of each image. Here we plot the mean signal from 1000 measurements to enhance the signal-to-noise ratio.}
\end{center}
\end{figure}

Importantly, the carrier frequency allows for immediate separation of all the MP-SPIFI images, circumventing the need to acquire multiple images to recover spatial frequency information beyond the cutoff of the imaging system. With appropriate selection of the rotational frequency of the mask, all images may be recovered in a single measurement.  Figure~\ref{fig:schematic}b shows the nonlinear illumination intensity in the lateral dimension versus scan time, as well as the corresponding decomposition of the intensity into harmonics of $f_{x,2}(t)$, where the carrier frequency for a given lateral position is seen to increase harmonically for each order. This is also clear from the experimental TPEF-SPIFI data displayed in Fig.~\ref{fig:signals}.

Inserting the nonlinear illumination intensity into Eq.~\eqref{eq:signal}, we find that the MP-SPIFI signal is a sum of $2 \, \eta$ images, each with lateral spatial frequency support up to $q \, f_{x,c}$ and carrier frequency $q \, \nu_c$:
\begin{equation}
S(t) = \sum \limits_{q = 0}^{2 \eta} b_q(t) \left< \cos \left[ 2 \pi \, q \, f_{x,2}(t) \, x + q \, \nu_c \, t \right] \, C(x,t) \right>_{\bf r} = \sum \limits_{q = 0}^{2 \eta} S_q(t)
\end{equation}
For each rotation of the modulation mask, the voltage vs.~scan time measurement of  $\beta({\bf r},t)$ from the photodetector encodes $2 \, \eta$ line images, all with varying spatial frequency support. The images are easily separated from one another by a Fourier transform of the photodetector signal, as the carrier frequency associated with each MP-SPIFI order ensures that the images do not overlap in the modulation frequency spectrum with proper selection of $\nu_r$.

Utilizing the Abbe definition we can write the spatial resolution of the $q^\mathrm{th}$ MP-SPIFI order as: $\delta x_q = \lambda/ (2 \, \mathrm{NA}_q) = 1/ (2 \, q \, f_{x,c})$, where $\mathrm{NA}_q$ is the effective NA of the $q^\mathrm{th}$-order image. Since the maximal spatial frequency support is achieved when $q = 2 \, \eta$, the best possible spatial resolution in MP-SPIFI is $\delta x _\mathrm{min} = \delta x_1/(2 \, \eta)$ when $j \in \left\{-1,0,1 \right\}$.

\section{Results}
To demonstrate that MP-SPIFI provides images with frequency support beyond the diffraction limit for contrast mechanisms with both real and virtual energy states, we used MP-SPIFI to image TPEF from 100-nm-diameter fluorescent nanodiamonds (FNDs), and SHG from 200-nm-diameter barium-titanate oxide (BaTiO$_3$) crystals. The objects were illuminated with femtosecond laser pulses centered at a wavelength of 1065~nm and focused by a 0.8~NA objective lens. By isolating and demodulating each MP-SPIFI order, the 1D optical transfer function (OTF) was recovered for each image. 

\begin{figure}
\begin{center}
\resizebox{0.5\linewidth}{!}{\includegraphics{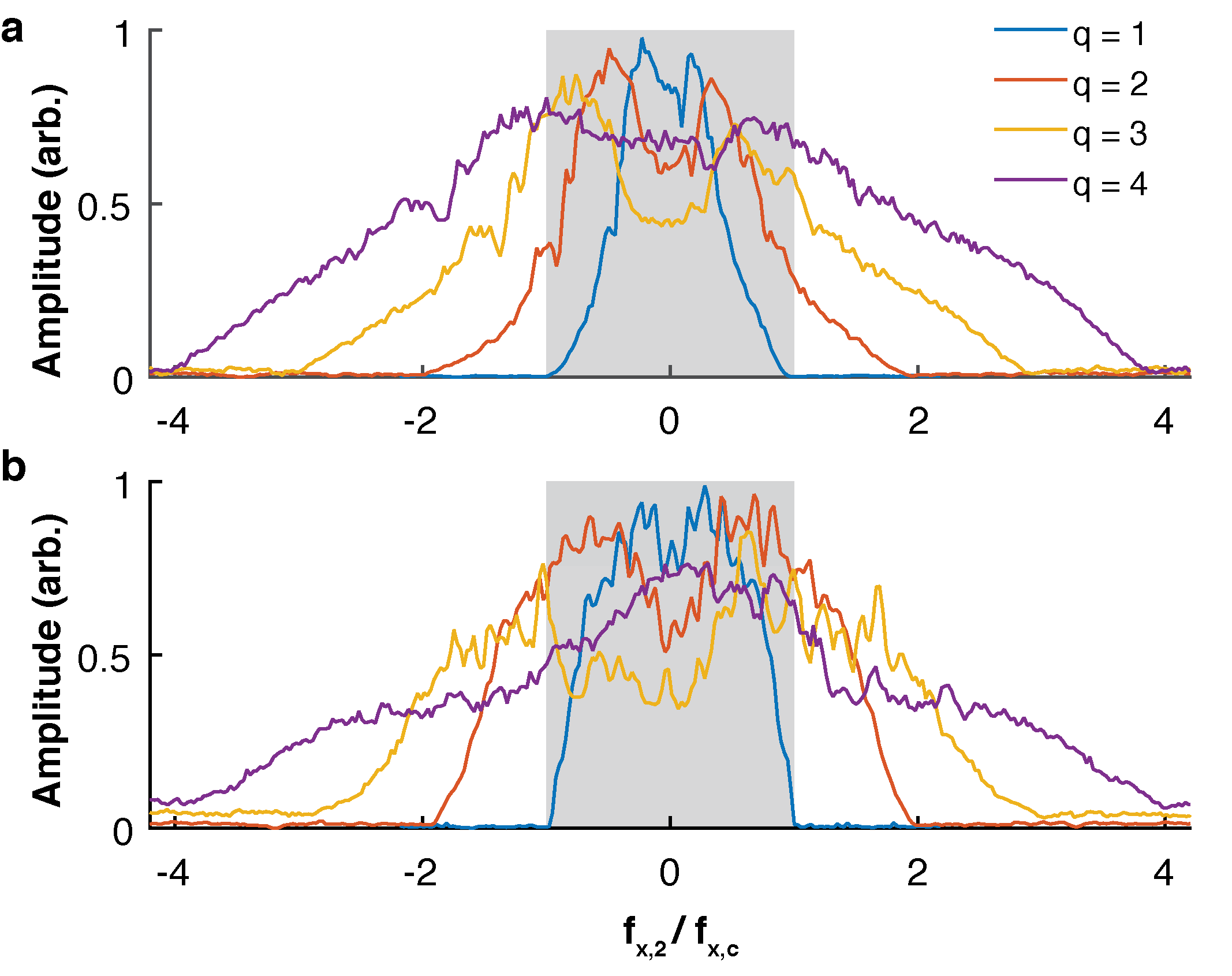}}
\caption{\label{fig:otfs} MP-SPIFI provides increased lateral spatial frequency support. Experimental 1D MTFs from (a) TPEF, measured from 100-nm FNDs, and (b) SHG, measured from 200-nm BaTiO$_3$ particles. The gray shaded regions represent the bounds of frequency support for linearly-excited imaging with $\lambda$=1065~nm and NA=0.8.}
\end{center}
\end{figure}

Figure~\ref{fig:otfs} shows the modulation transfer function (MTF) for each SPIFI order, $\left| S_q(f_{x,2}) \right|$, where we have used the fact that $f_{x,2}(t) = (n_2/n_1) \, M \, \Delta k \, \nu_r \, t$ to scale the temporal axis to the lateral spatial frequency domain. It is clear that lateral spatial frequency information beyond the cutoff frequency of the objective lens is contained in all MP-SPIFI orders for $q>1$.

\begin{figure*}
\begin{center}
\resizebox{\linewidth}{!}{\includegraphics{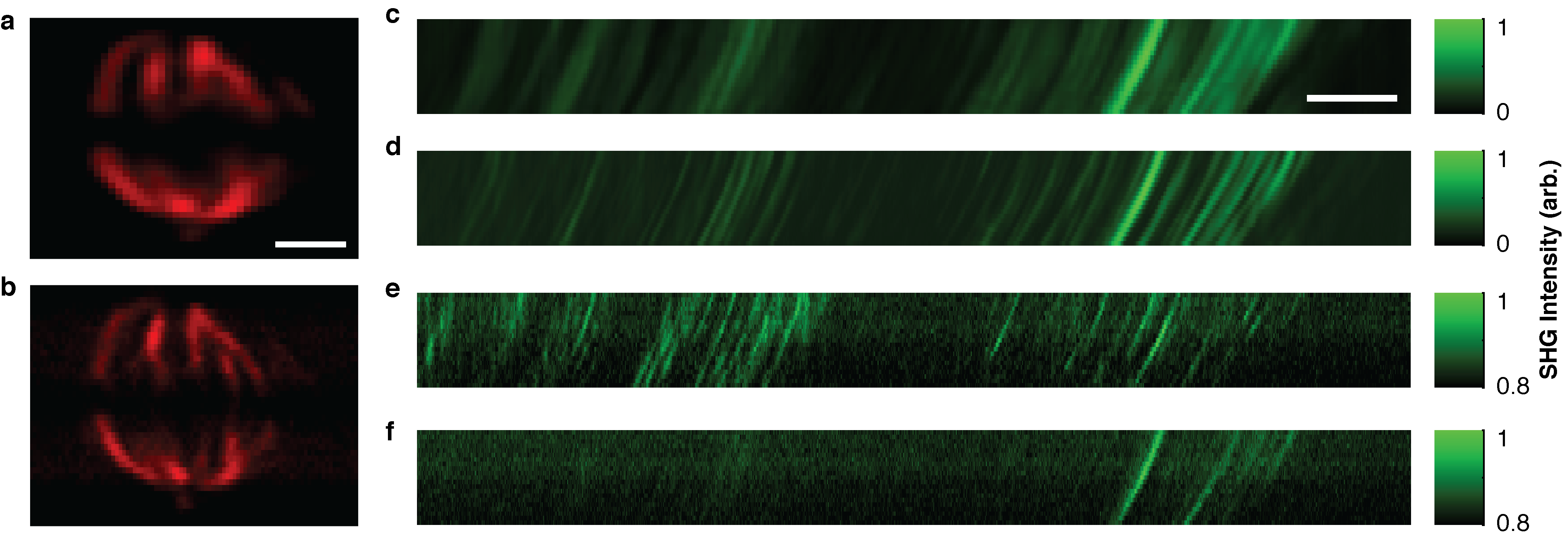}}
\caption{\label{fig:biology}TPEF and SHG images collected from biological media with MP-SPIFI. (a) First- and  (b) second-order MP-SPIFI images of TPEF from a mitotic HeLa cell immunostained with primary antibodies against alpha tubulin and secondary antibodies tagged with Alexa 546. Scale bar:  3~$\mu$m. (c)--(f) First through fourth order images of SHG from fixed, 16-$\mu$m-thick rabbit tendon. Ringing in the third-order image (e) was attributed to modulations in the corresponding MTF (cf. Fig.~\ref{fig:otfs}). Scale bar: 10~$\mu$m. All images were collected at 0.8~NA with laser pulses centered at 1065~nm. TPEF was collected in the epi direction and SHG was collected in the forward-scattered direction.}
\end{center}
\end{figure*}

We demonstrated super-resolved multiphoton imaging of biological media by imaging TPEF from polymerized tubulin fibers (i.e.~microtubules) tagged with with Alexa 546 in a mitotic HeLa cell, and SHG from collagen fibers in fixed rabbit tendon (Fig.~\ref{fig:biology}). One-dimensional MP-SPIFI images were collected in the horizontal dimension as the specimen was scanned vertically to form a two-dimensional (2D) image. Resolution enhancement occurs only in the horizontal dimension. Lateral tomography has been used for 2D image capture with linear SPIFI, and could be applied to MP-SPIFI for super-resolved 2D imaging \cite{Schlup:2011}.

\begin{figure*}
\begin{center}
\resizebox{\linewidth}{!}{\includegraphics{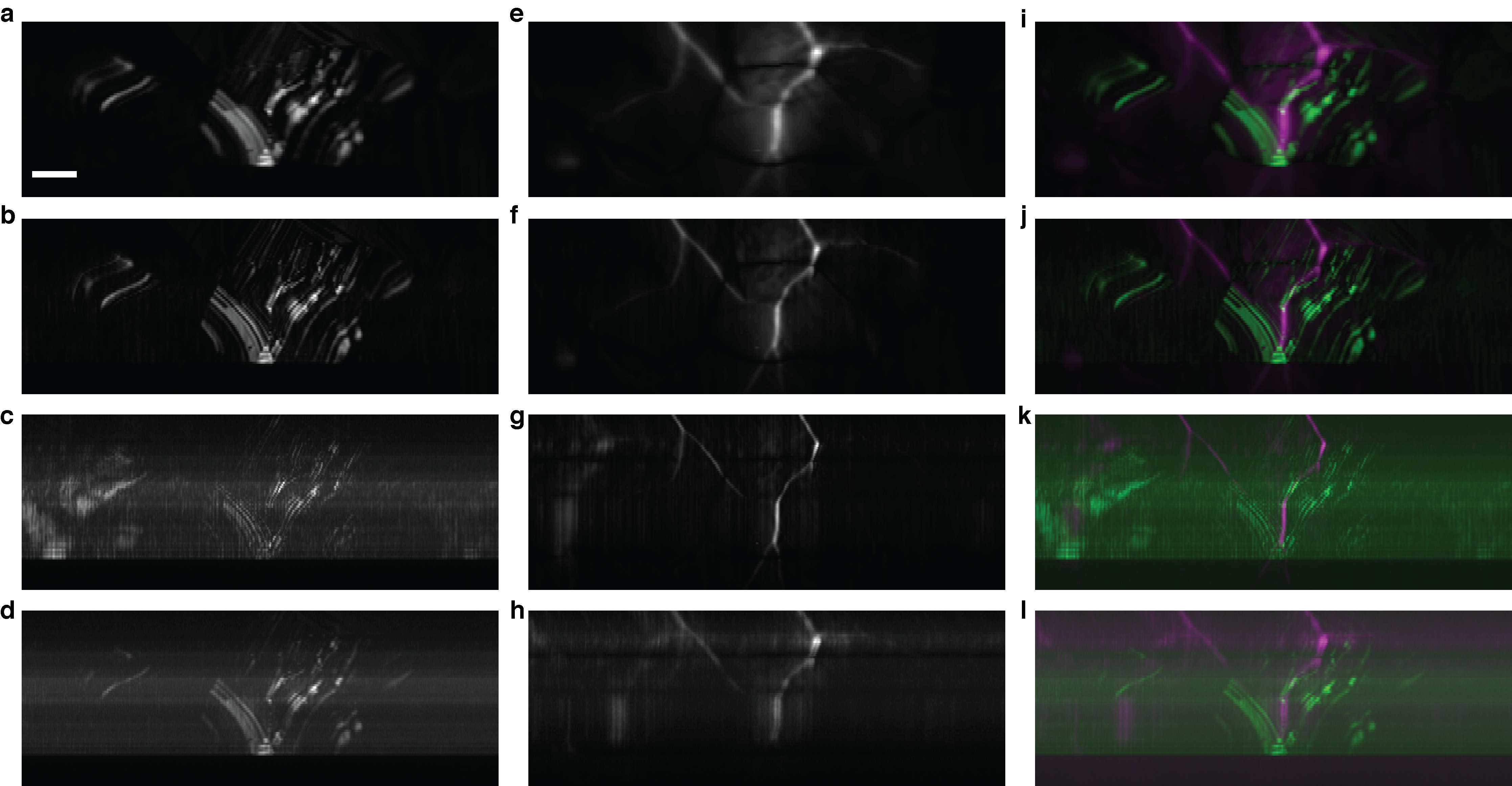}}
\caption{\label{fig:cdte} Multimodal super-resolved images of SHG and PL from CdTe solar cells. (a)--(d) First through fourth order SHG images. (e)--(h) Corresponding  PL images. (i)--(l) False color merged images. PL and SHG were measured in the epi-direction simultaneously. Scale bar: 10~$\mu$m.}
\end{center}
\end{figure*}

To demonstrate simultaneous multimodal super-resolved imaging, we imaged photoluminescence (PL) and SHG from CdTe solar cells (Fig.~\ref{fig:cdte}). Due to the opacity of the CdTe cells, both contrast mechanisms were collected simultaneously in the epi-direction. Figures~\ref{fig:cdte}a--d display the first- through fourth-order SHG-SPIFI images, and the corresponding PL-SPIFI images are shown in Fig.~\ref{fig:cdte}e--h. Comparison of the images for each MP-SPIFI order makes it clear that the spatial resolution is enhanced in both SHG and PL simultaneously, although the resolution appears to be different for these two contrast modes beyond the second-order data. The reason for this discrepancy is unclear at this time, though we note that the relatively long lifetime of the PL contrast mechanism, as well as diffusion of excited carriers may be responsible for these differences.

\section{Discussion}
Although the spatial frequency support attainable for TPEF and SHG imaging with MP-SPIFI is four times that provided by the objective lens, not all images shown in this work reach this theoretical limit. For example, TPEF-SPIFI images obtained from tubulin in HeLa cells were limited to a resolution enhancement of 2$\times$. The loss of spatial frequency information in higher-order MP-SPIFI images is attributed to the reduction in signal-to-noise ratio (SNR) with increasing MP-SPIFI order. This loss of SNR has two primary causes.

First, the noise floor in the modulation frequency (spatial) domain is determined by the shot noise from the time-averaged total light intensity measured on the photodetector, $\overline{\dirac{\beta({\bf r},t)}_{\bf r}}$. Conversely, each MP-SPIFI order represents only a portion of this total light intensity. Since the shot-noise-limited noise floor  scales as $\overline{\dirac{\beta({\bf r},t)}_{\bf r}}$, the SNR of higher order MP-SPIFI images scales as $\dirac{\beta_{q}({\bf r},t)}_{\bf r}/\overline{\dirac{\beta({\bf r},t)}_{\bf r}}$. This has also been observed in fluorescent imaging using radiofrequency-tagged emission (FIRE) \cite{Diebold:2013}, another imaging method that uses spatial frequency-modulated illumination to form images, and the spatial analog has been analyzed in detail for    off-axis SHG holography \cite{Smith:2013}. Images of TPEF collected from HeLa cells truncated at second-order due to a large noise floor that restricted our ability to collect higher-order images. On the other hand, images of SHG from rabbit tendon had a significantly larger SNR due to increasing the power of the illumination laser by $\sim$50\%, making the fourth-order signal detectible above the shot noise floor. 

A possible route to enhance the SNR in higher-order MP-SPIFI images is the use of a phase modulator in place of the amplitude modulating disk, thereby eliminating the undiffracted ($j=0$) beam in the object region and significantly reducing the shot noise floor.

Secondly, vignetting of the diffracted beams ($|j| > 0$) in the pupil plane of the objective lens causes a reduction in both transmitted energy and spatial frequency support in the vertical ($y$) dimension. We investigated the effect of the circular objective pupil for each MP-SPIFI order using numerical simulations. First, the throughput energy of each diffracted order with scan time was computed with a simple overlap integral (Fig.~S\ref{sfig3}). The loss in energy of the higher-order diffracted beams, particularly the $j = \pm1$ beams, causes a reduction in the modulation depth of the illumination intensity pattern, reducing the amplitude of each MP-SPIFI order with respect to the background DC signal. 

We also investigated the effect of the circular pupil aperture on the spatial frequency support of the diffracted beams in the vertical dimension with a numeric calculation of the electric fields in the object region with a vectorial angular spectrum formalism \cite{Born:1999,Novotny:2012} (Supplementary Information). Comparison of the spatial frequency support obtained from the numeric model confirms that the spatial frequency support is reduced by vignetting from the objective lens (Fig.~S\ref{fig:scan_summary}). Despite this effect, MP-SPIFI shows promise for attaining multimodal super-resolved images of nonlinearly-excited contrast mechanisms.  

\begin{figure}
\begin{center}
\resizebox{0.5\linewidth}{!}{\includegraphics{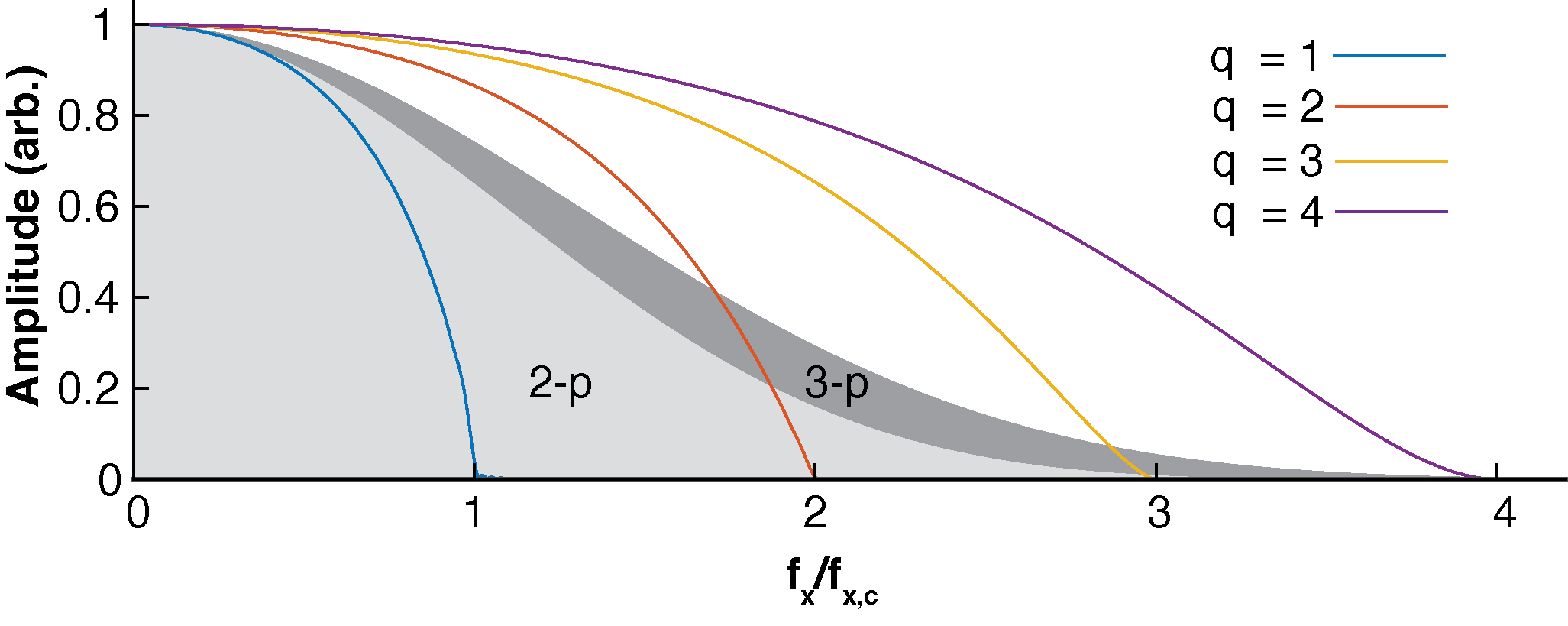}}
\caption{\label{fig:tplsm_vs_tpspifi} Comparison of the amplitude of the transfer function in multiphoton laser scanning microscopy and two-photon SPIFI. The shaded grey regions indicate the frequency support for two- and three-photon laser scanning microscopy. Solid lines indicate the frequency support in TP-SPIFI where only first-order diffraction from the modulation mask was included. All OTFs were computed with the Debye integral using $\lambda = 1065~\mathrm{nm}$ and 0.8 NA.}
\end{center}
\end{figure}

The ability to resolve fine spatial features in an image depends on both the shape and the extent of the spatial frequency support of the OTF. To evaluate the relative imaging performance of MP-SPIFI compared to conventional LSM techniques, the OTFs of these methods were computed using the vector focusing numerical model for two-photon SPIFI (TP-SPIFI), and two- and three-photon LSM. Intensities were computed for an illumination wavelength of 1065~nm, 0.8 NA, and linearly-polarized light aligned with the vertical ($y$) direction. For TP-SPIFI, the widths of the intensity distributions in the pupil plane were modeled after the MP-SPIFI system reported here (Supplementary Information). For the multiphoton LSM calculation, the beam size in the vertical dimension of the TP-SPIFI microscope was used as the radius of the circularly-symmetric Gaussian distribution. The lateral spatial frequency support was then computed from the illumination intensity for each imaging type. The resulting spatial frequency support is shown in Fig.~\ref{fig:tplsm_vs_tpspifi}, where the shaded grey regions indicate the spatial frequency support of each LSM mode, while solid lines indicate the spatial frequency support for each TP-SPIFI order. 

What is clear in Fig.~\ref{fig:tplsm_vs_tpspifi} is that MP-SPIFI imaging collects significantly more information content at high spatial frequencies than multiphoton LSM, thus providing the ability to extract significantly more spatial information. As the scan time progresses in TP-SPIFI, the focusing of the transmitted diffracted orders is perturbed by the circular objective pupil, which shapes the spatial frequency support of the OTF. The MTFs associated with TP-SPIFI extend to greater spatial frequency extent for the second-order nonlinear interaction than either two- or three-photon LSM. Moreover, the amplitude of the high spatial frequency content for TP-SPIFI is significantly higher than for multiphoton LSM. The result is that TP-SPIFI captures significantly more fine spatial information about an image that multiphoton LSM.

In summary, we have described a new approach to super-resolved imaging in multiphoton excited microscopy that is capable of measuring spatial frequency information beyond the diffraction limit for both multiphoton fluorescence and coherent scattering. The combination of excitation wavelengths in the NIR and single-pixel detection should allow MP-SPIFI to form images with enhanced spatial resolution at penetration depths well beyond those attainable with linearly-excited fluorescence. MP-SPIFI enables super-resolved imaging of new contrast mechanisms, which may lead to new insights in both biological and material science studies.

\section{Materials and Methods}
\noindent{\bf Femtosecond laser source.} The microscope system is illuminated by pulses from a nonlinear fiber amplifier centered at 1065~nm, built in-house \cite{Domingue:2014}.  The pulses are generated by an all-normal dispersion fiber oscillator using Yb-doped gain fiber \cite{Chong:2006}, passed through a tunable narrow spectral bandpass filter, and then amplified in Yb-doped fiber.  The tunable filter varies the launch condition into the nonlinear amplifier, resulting in varying spectral broadening via self-phase-modulation.  At our typical operating point this resulted in a bandwidth of 28~nm (full-width at half-maximum; FWHM).  The pulses are temporally compressed in a folded Martinez compressor, resulting in near transform-limited durations of approximately 150~fs at an average power ranging from 900~mW to 1.3~W at a repetition rate of 52.5~MHz.

\vspace{5pt}
\noindent{\bf Optical setup.} The beam from the laser is collimated and expanded with a 4:1 Keplerian telescope composed of two achromatic lenses (ThorLabs, AC254-100-C-ML and AC254-400-C-ML) obtain a beam size of 8.85~mm (full-width at half-maximum of the intensity).  The beam is brought to a line focus on the modulation mask with an achromatic cylindrical lens (ThorLabs, ACY-254-150-B) oriented such that the beam is focused in the vertical ($y$) direction only. 

The mask plane is re-imaged to the object plane in two stages, first with a 1:1 image relay system composed of two achromatic lenses (ThorLabs AC254-100-C-ML) in a 4-$f$ configuration, and again with an image relay system consisting of a tube lens (ThorLabs AC254-150-C-ML) and an infinity-corrected objective lens. Data presented in this work were collected with a 0.8~NA objective lens (Zeiss, N-Achroplan 50x/0.8 NA Pol), such that the mask is imaged to the object region with de-magnification of 46.5.

\vspace{5pt}
\noindent{\bf Modulation mask.} The modulation mask consists of a pattern defined in polar coordinates by the expression $m(r,\varphi) = 1/2 + 1/2 \,  \mathrm{sgn}[\cos(\Delta k \, r \, \varphi)]$. The parameter $\Delta k$ defines the density of features on the mask.  The modulation mask used in this work has a density parameter $\Delta k$ = 70/mm. The disk was mounted on a motor (Faulhaber, 2057S012BK1155) with a custom chuck. The rotation speed of the disk was controlled by an external speed controller (Faulhaber, MCBL3006). 

While the modulation mask imparts transverse spatial frequencies to the line focus in both the $x$ and $y$ dimensions, only the spatial frequency in the $x$ direction varies with the rotation angle of the mask.  The incident beam is diffracted into multiple orders denoted by the integer $j$. Since the modulation mask is composed of a binary amplitude modulation pattern, the set of diffracted orders for an ideal mask is composed of only the undiffracted beam ($j$ = 0) and odd diffracted orders. 

\vspace{5pt}
\noindent{\bf Signal collection and digitization.} All images displayed in this work were collected in the epi-direction, which was achieved by use of a long-pass dichroic beamsplitter placed between the tube lens and the objective lens (Semrock, FF875-Di01).

Fluorescent light from the HeLa specimen was collected in the epi-direction through a short-pass filter (Semrock, FF720/SP) and a 500 nm long-pass filter (Chroma, ET500lp). 

\vspace{5pt}
\noindent{\bf Sub-diffraction-limited objects.} Samples used for point spread function measurements were 200~nm BaTi0$_3$ particles (1148DY, Nanostructured \& Amorphous Materials, Houston TX) for SHG, and 100~nm fluorescent nanodiamonds (ND-400NV-100nm-10mL, Adamas Nanotechnologies, Raleigh NC) for TPEF.  In both cases the insoluble nanoparticles, suspended in water, were diluted, placed in an ultrasonic bath to break up aggregated particles, and then drop cast onto a microscope slide, where the water evaporated away before being cover-slipped. 

\vspace{5pt}
\noindent{\bf Immunofluorescence.} HeLa cells were seeded onto glass coverslips and fixed when they reached $\sim$70\% confluency. Cells were rinsed rapidly with 37~C 1X PHEM buffer (60~mM PIPES, 25~mM HEPES, 10~mM EGTA, 4~mM MgSO$_4$, pH 7.0) followed by lysis at 37~C for 5~min in freshly prepared lysis buffer (1X PHEM + 0.5\% Triton-X-100). Cells were then fixed on the bench top for 3~min using ice cold methanol (95\% methanol + 5~mM EGTA) followed by an additional 20~min methanol fixation at 20~C. At room temperature, cells were then rehydrated with 1X PHEM then rinsed 3$\times$5 min in PHEM-T (1X PHEM + 0.1\% Triton-X-100) and blocked in 10\% boiled donkey serum (BDS) in PHEM for 1~hr at room temperature. Microtubules were labeled with anti-alpha tubulin primary antibodies (Sigma-Aldrich, St. Louis, MO) diluted in 5\% BDS for 12~hr at 4~C. Following primary antibody incubation, cells were rinsed 3$\times$5 min in PHEM-T and then incubated for 45 min at room temperature with secondary antibodies conjugated to Alexa 546 (Life Technologies). Cells were then rinsed 3$\times$5 min in PHEM-T, incubated in a solution of 2~ng/mL 4,6-diamidino-2-phenylindole (DAPI) diluted in PHEM, rinsed again (3$\times$5 minutes), and then mounted onto microscope slides in an anti-fade solution containing 90\% glycerol and 0.5\% N-propyl gallate.  Coverslips were sealed with nail polish to affix them to the slides. 

\vspace{5pt}
\noindent{\bf CdTe crystals.} A polycrystalline, large grained CdTe sample was grown by close spaced sublimation (CSS), treated with CdCl$_2$, and polished. This has the effect of ``sealing'' small grains ($\sim$2~$\mu$m) after growth such that average grain size becomes larger. The sample was polished using an ion beam technique, basically like sputtering yet instead of adsorption, it simply erodes irregularities in structure away such that the surface is smooth.

\vspace{5pt}
\noindent{\bf Data analysis.} All data analysis was performed in MATLAB (The MathWords, Natick, NJ).  Each 1D line image was reconstructed by computing the fast Fourier transform (FFT) of the measured temporal data $S(t)$ to obtain the MP-SPIFI images in the modulation frequency domain, i.e., $\tilde{S}(\nu_t)$. Since $S(t)$ is a real signal, copies of each image are located at both positive and negative carrier frequencies in the modulation frequency domain. Each MP-SPIFI order was recovered by numerically filtering at the positive carrier frequency. The filtered data, $\tilde{S}_{q}(\nu_t)$, was inverse transformed to obtain $S_{q+}(t)$. The complex temporal trace was then demodulated by the relative carrier frequency, $q \, \nu_c$, and the temporal axis was calibrated according to the relation $f_{x,2}(t) = q \, (n_2/n_1) \, M \, \Delta k \, \nu_r \, t$ to obtain $S_{q+}(f_{x,2})$. Images were obtained by again taking the FFT of the data.

Data was calibrated by measuring light transmitted through a 2~$\mu$m slit in the object plane as a function of lateral position. The centroid of the measured distribution was computed as a function of lateral position.  This allowed the quantity describing the relationship between lateral position and carrier frequency, $\kappa = (n_2/n_1) \, M \, \Delta k \, \nu_r$ to be recovered empirically.

Imperfections in disk mounting lead to an additional phase accumulated in $S(t)$ as a function of mask rotation. The process used to remove disk aberration phase has been discussed at length in a previous study \cite{Field:2015:01}

\section*{Acknowledgments}
JJF, KAW, SRD and RAB acknowledge funding from the W.M. Keck Foundation. JAS acknowledges support from the National Institute of Biomedical Imaging and Bioengineering under the Bioengineering Research Partnership EB-00382. We thank James Burst for growing the CdTe specimens and Susanta Sarkar for donating the fluorescent nanodiamonds.

\section*{Author contributions}
JJF and RAB designed the experiments. SRD and RAB designed the femtosecond laser source. SRD and KAW constructed the laser source. JJF designed the MP-SPIFI microscope, and JJF and KAW constructed the microscope. JJF and KAW collected the data. AAM assisted in collecting data from CdTe crystals. JJF and KAW performed the data analysis. JJF and RAB developed the theoretical framework. JJF carried out the numeric simulations. KFD and JGD provided HeLa cells. AAM, DHL, DK and JAS provided CdTe samples. JJF, KAW, JAS and RAB interpreted the data. JJF and RAB wrote the manuscript. All authors contributed to editing the final manuscript.

\section*{Supplementary Information}

\setcounter{figure}{0}
\makeatletter
\makeatletter \renewcommand{\fnum@figure}
{\figurename~S\thefigure}
\makeatother

\section{Theoretical analysis of super-resolved MP-SPIFI}

Here we derive the mathematical framework for MP-SPIFI and show how spatial frequency information that lies outside the cutoff frequency of the objective lens is encoded into a temporal measurement of signal light with a single-element photodetector. We begin by deriving a general expression for the illumination intensity, and hence the signal, then consider the case where only first-order diffraction from the modulation mask contributes to the signal.

Throughout the analytic derivation, we consider plane wave illumination of the modulation mask for simplicity.  We have rigorously analyzed plane wave illumination in a previous report and adopt much of the notation here \cite{Field:2015:02}.

\subsection{General expressions}
\begin{figure}[!ht]
\begin{center}
\resizebox{\linewidth}{!}{\includegraphics{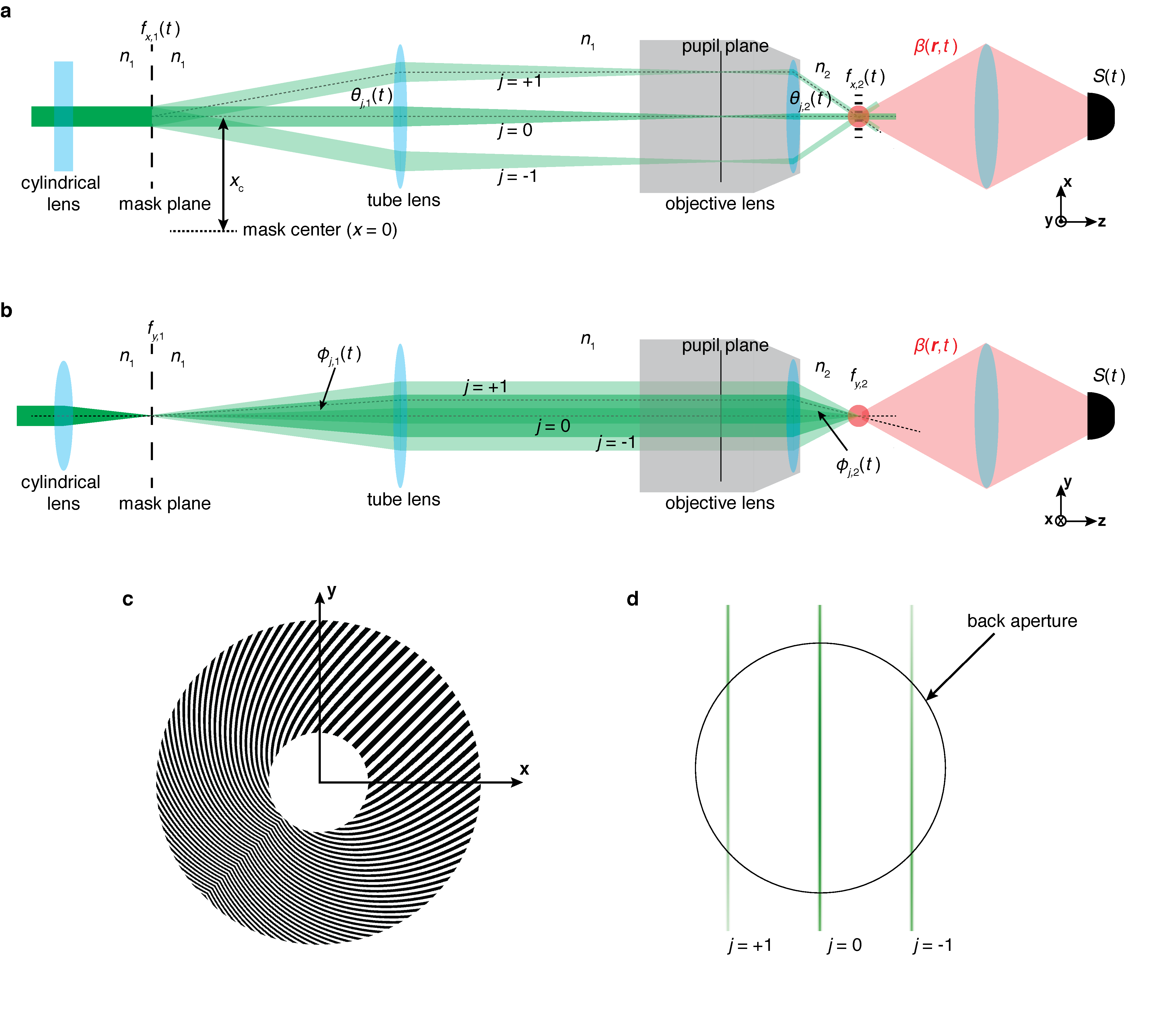}}
\caption{\label{sfig1} MP-SPIFI system configuration. System schematics in the (a) $(x,z)$ and (b) $(y,z)$ planes. (c) The SPIFI modulation mask with $\Delta k$ = 2/mm. (d) The spatial distribution of diffracted beams in the pupil plane, computed using the parameters of the imaging system used in this work ($\lambda$=1064~nm, NA=0.8, tube lens focal length of 150~mm). The back aperture diameter was computed using the magnification of the objective lens, $M_o$, and the design tube lens focal length $F_t'$=164.5~mm (Zeiss UIS series) with the expression: $d = 2 \, F_t' \, \mathrm{NA} / M_o$.}
\end{center}
\end{figure}

As discussed in the manuscript, images are formed in the SPIFI microscope by projecting illumination intensity patterns into the specimen and collecting signal light emitted from the object on a single-element detector. The illumination intensity is formed by the spatial interference of multiple light sheets propagating at varied angles with respect to the optic axis. The density (spatial frequency) of the interference fringes in the illumination intensity is precisely controlled by use of a circular glass disk on which is printed an amplitude modulation mask. A spatially-coherent illumination beam is brought to a line focus on the modulation mask, and the mask plane is then imaged to the object region with a tube lens and an objective lens (cf. Fig.~S\ref{sfig1}a and Fig.~S\ref{sfig1}b). 

During a single rotation of the disk, the fundamental spatial frequency of the interference fringes is varied through the entire range of spatial frequencies supported by the image relay system. As the disk rotates, the signal light, which is generated by the interaction of the illumination intensity and the distribution of molecules in the object, is collected on a single element detector. The temporal signal measured from the photodetector is expressed by integrating the contrast signal over all space. The contrast signal, $\beta({\bf r},t)$, is the product of the illumination intensity pattern raised to the power of the nonlinearity, $\eta$, and the contrast distribution function, $C({\bf r},t)$, which describes how the contrast-producing molecules within the object are spatially distributed as a function of time:
\begin{equation}
S(t) = \int \limits_{-\infty}^{\infty} \mathrm{d}^3 {\bf r} \, \, \beta({\bf r},t)  = \dirac{ \left[ I_\mathrm{ill}({\bf r},t) \right]^{\eta} \, C({\bf r},t) }_{\bf r}
\label{spifi_signal}
\end{equation}
We note that the nonlinear dependence on the illumination intensity need not be integer valued, as is the case in harmonic generation (HG) microscopy, for example. If the nonlinearity is not a simple power dependence, one could express the contrast light as $\beta({\bf r},t) = f \left[ I_\mathrm{ill}({\bf r},t) \right] \, C({\bf r},t)$, where $f[\cdot]$ is a function describing, for example, saturation of electronic absorption in target molecules \cite{Heintzmann:2009}. The functional dependence of the illumination intensity can be expanded in a Taylor series, and the highest power of the Taylor series that contributes to the signal will determine the cutoff spatial frequencies in super-resolved MP-SPIFI. To simplify the analysis that follows, we only consider the case where the contrast intensity depends on an integer power of the illumination intensity.

The signal measured from the photodetector, $S(t)$, represents the projection of the illumination intensity onto the spatial frequency content of the object.   We denote the lateral spatial frequency experienced by the illuminating light sheet in the mask plane as $f_{x,1}(t)$, while $f_{x,2}(t)$ is the spatial frequency in the object region. At a given scan time, $t_s$, the principle spatial frequency projected onto the object is $f_{x,2}(t_s)$, and correspondingly the signal amplitude of the signal $S(t_s)$ encodes the relative amplitude of $f_{x,2}(t_s)$ in the object. The whole temporal signal $S(t)$ represents the product of the spatial frequency distribution of the object and the spatial frequencies projected into the specimen, thus encoding a one-dimensional (1D) image of the object in the spatial frequency domain.

Here we wish to derive the form of the signal from the photodetector, $S(t)$. Since the illumination beam is spatially coherent, the modulation mask causes the incident light beam to be diffracted into multiple beams propagating at differing angles with respect to the optic axis. Each diffracted beam has a corresponding  wavevector ${\bf k}_{j,1}(t) = k_{x,j,1}(t) \, {\bf x} + k_{y,j,1}(t) \, {\bf y} + k_{z,j,1}(t) \, {\bf z}$.  Directly behind the mask, we may write the spatial portion of the electric field for the diffracted order $j$ as:
\begin{equation}
v_{j,1}({\bf r},t) = a_j \, \ee^{\ii \, {\bf k}_{j,1}(t) \cdot {\bf r}} =  a_j \, \ee^{\ii \, k_{x,j,1}(t) \, x} \ee^{\ii \, k_{y,j,1}(t) \, y} \, \ee^{\ii \, k_{z,j,1}(t) \, z}
\end{equation}
where $k_{z,j,1}(t) = \sqrt{k_1^2 - k_{x,j,1}^2(t) - k_{y,j,1}^2(t)}$, and $k_1 = 2 \pi \, n_1 / \lambda$ is the wavenumber of the illumination beam in the mask region. 

To obtain the electric field in the object region, we apply the magnification of the imaging system to the transverse wavevectors of the electric field in the mask region. Specifically, the transverse wavevectors in the object region are:
\begin{eqnarray}
k_{x,j,2}(t) & = & \frac{n_2}{n_1} \, M \, k_{x,j,1}(t) \\
k_{y,j,2}(t) & = & \frac{n_2}{n_1} \, M \, k_{y,j,1}(t)
\end{eqnarray}
so we may write the electric field in the object region as:
\begin{equation}
v_{j,2}({\bf r},t) = a_j \, \ee^{\ii \, {\bf k}_{j,2}(t) \cdot {\bf r}} =  a_j \, \exp \left[\ii \, \frac{n_2}{n_1} \, M \, k_{x,j,1}(t) \, x \right] \exp \left[ \ii \, \frac{n_2}{n_1} \, M \, k_{y,j,1}(t) \, y \right] \, \exp \left[ \ii \, k_{z,j,2}(t) \, z \right]
\end{equation}
where $k_{z,j,2}(t) = \sqrt{k_2^2 - \left[ k_{x,j,2}(t) \right]^2 - \left[ k_{y,j,2}(t)\right]^2 }$, and $k_2 = 2 \pi \, n_2 / \lambda$ is the wavenumber in the object region.

The illuminating electric field in the object plane is a sum of the electric fields in the object region:
\begin{equation}
E_\mathrm{ill}({\bf r},t) = \sum \limits_{j = -N}^{N} v_{j,2}({\bf r},t)
\end{equation}
Defining the conjugate of the $j^{\mathrm{th}}$ diffracted beam as $u_{j,2}({\bf r},t) \equiv \left[v_{j,2}({\bf r},t)\right]^{*}$, we can write the nonlinear intensity as:
\begin{equation}
\left[ I_\mathrm{ill}({\bf r},t) \right]^\eta = \left[ \sum \limits_{j = -N}^{N} \sum \limits_{k = -N}^{N} v_{j,2}({\bf r},t) \,  u_{k,2}({\bf r},t)\right]^\eta
\end{equation}
Finally, the signal from the photodetector is:
\begin{equation}
S(t) = \dirac{\left[ \sum \limits_{j = -N}^{N} \sum \limits_{k = -N}^{N} v_{j,2}({\bf r},t) \,  u_{k,2}({\bf r},t)\right]^\eta \, C({\bf r},t) }_{\bf r}
\end{equation}

\subsection{First-order diffraction only}
While the amplitude modulation masks used in this work are binary, and thus have numerous diffracted orders occurring at odd-order harmonics, it is instructive to consider amplitude masks for which only first-order diffraction occurs. Formally, this is equivalent to considering an amplitude modulation mask for which the printing process is not binary, but instead can faithfully produce sinusoidal amplitude modulation. Using the results of \cite{Field:2015:02}, the electric field for each beam in the object region is:
\begin{eqnarray}
v_{0,2}({\bf r},t) & = & a_0 \, \exp \left( \ii \, k_2 \, z \right) \\
v_{1,2}({\bf r},t) & = & a_1 \, w_1(t) \, \exp \left[\ii \, k_{x,1,2}(t) \, x \right] \exp \left[\ii \, k_{y,1,2}(t) \, y \right] \exp \left[\ii \, k_{z,1,2}(t) \, z \right] \\
v_{-1,2}({\bf r},t) & = & a_1 \, w_{-1}(t) \, \exp \left[\ii \, k_{x,-1,2}(t) \, x \right] \exp \left[\ii \, k_{y,-1,2}(t) \, y \right] \exp \left[\ii \, k_{z,-1,2}(t) \, z \right]
\end{eqnarray}
where $w_j(t)$ is the temporal window function, which accounts for time-limited diffraction from the modulation mask and apodization of the first-order diffracted beams as they scan across the back aperture of the objective lens. For a well-aligned system, the window functions are symmetric about time zero, so we can write $w(t) = w_{-1}(t) = w_1(t)$. Note that we have assumed that $w(t)$ is a real-valued function, and that the amplitude $a_1$ is a real constant. 

The angles in the object region $\theta_{1,2}(t)$ and $\phi_{1,2}(t)$ are related to the spatial frequencies in the transverse dimensions. From Fig.~S\ref{sfig1} it is clear that the angles of conjugate diffracted beams, e.g., $j = \pm1$, display symmetry about the optic axis. Consequently, we can express the lateral and vertical spatial frequencies of the first-order diffracted beams in the object region as:
\begin{eqnarray}
k_{x,\pm1,2}(t) & = &  k_2 \, \sin \theta_{\pm1,2}(t) = \pm k_2 \, \sin \theta_{1,2}(t) \\
k_{y,\pm1,2}(t) & = &  k_2 \, \sin \phi_{\pm1,2}(t) = \pm k_2 \, \sin \phi_{1,2}(t) 
\end{eqnarray}
From these expressions is is clear that $k_{x,-1,2}(t) = - k_{x,1,2}(t)$ and $k_{y,-1,2}(t) = - k_{y,1,2}(t)$. Using these expressions, we also find that $k_{z,1,2}(t) = k_{z,-1,2}(t)$. Using these simplifications, the electric field of the $j = -1$ beam in the object region becomes:
\begin{equation}
v_{-1,2}({\bf r},t) = a_1 \, w(t) \, \exp \left[-\ii \, k_{x,1,2}(t) \, x \right] \exp \left[-\ii \, k_{y,1,2}(t) \, y \right] \exp \left[\ii \, k_{z,1,2}(t) \, z \right] 
\end{equation}
Thus we can express the total electric field in the object region as:
\begin{equation}
E_\mathrm{ill}({\bf r},t) = \sum \limits_{j = -1}^{1} v_{j,2}({\bf r},t)  = a_0 \ee^{\ii \, k_2 \, z} + 2 \,a_1 \, w(t) \, \cos \left[ k_{x,1,2}(t) \, x + k_{y,1,2} \, y \right] \, \exp \left[ \ii \, k_{z,1,2}(t) \, z \right] 
\end{equation}

Since the illumination beams are tightly focused in the vertical dimension by the objective lens, nonlinear excitation results in an optically-sectioned image at the focal plane of the objective lens. In the remainder of this analysis, we therefore set $z = 0$.  We also choose to set $y =0$ to examine the modulation intensity along the optic axis of the system. Finally, we assume that the amplitudes can be set by considering the diffraction efficiency from a perfect square grating, such that $a_0 = 1/2$ and $a_1 = 1/\pi$. Applying this set of simplifications to the electric field distribution above, the illumination intensity for a second-order nonlinear process ($\eta = 2$) becomes:
\begin{eqnarray}
I_\mathrm{ill}^2(x,t) & = & \left[\frac{1}{16} +  \frac{3}{\pi^2}\, w^2(t) + \frac{6}{\pi^4} \, w^4(t)  \right] + \left[ \frac{1}{\pi} \, w(t) + \frac{12}{\pi^3} \, w^3(t) \right] \, \cos \left[ k_{x,2}(t) \, x + \nu_c \, t \right] \nonumber \\
	& & + \, \left[ \frac{3}{\pi^2} \, w^2(t) + \frac{8}{\pi^4} \, w^4(t) \right] \, \cos \left[ 2 k_{x,2}(t) \, x + 2 \nu_c \, t\right]  + \, \frac{4}{\pi^3} \, w^3(t) \, \cos \left[ 3 k_{x,2}(t) \, x + 3 \nu_c \, t \right] \nonumber \\
	& & \, \frac{2}{\pi^4} \, w^4(t) \, \cos \left[ 4 k_{x,2}(t) \, x + 4 \nu_c \, t \right]
\end{eqnarray}

In the preceding expression, we have explicitly written in the carrier frequency, $\nu_c$, which results from the optic axis being aligned at a lateral position $x_c \neq 0$, as shown in Fig.~S\ref{sfig1}a. The lateral position relative to the center of the modulation pattern can be written as $x' = x_c +  x$, where $x$ is the lateral position relative to the optic axis. The product of the spatial frequency in the object region and the lateral position $x'$ becomes:
\begin{equation}
k_{x,1,2}(t) \, x' = k_{x,1,2}(t) \, x_c + k_{x,1,2}(t) \, x = 2 \pi \, \nu_c \, t + k_{x,1,2}(t) \, x
\end{equation}
The angular spatial frequencies in the object region can be related to the scan time by considering the angles of each diffracted beam with respect to the optic axis. We have previously computed these angular frequencies to be \cite{Field:2015:02}:
\begin{equation}
k_{x,1,2}(t)  =  2 \pi \, f_{x,1,2}(t)  = 2 \pi \, \frac{n_2}{n_1} \, M \, f_{x,1,1}(t) = 2 \pi \, \frac{n_2}{n_1} \, M \, \Delta k \, \nu_r \, t 
\end{equation}
The carrier frequency can then be written as $\nu_c = (n_2/n_1) \, M \, \Delta k \, \nu_r \, x_c$.

This analysis can be generalized for an arbitrary integer nonlinearity to find:
\begin{equation}
\left[ I_\mathrm{ill}(x,t) \right]^\eta = \sum \limits_{q = 0}^{2 \eta} b_q(t) \, \cos \left[ q \, k_{x,1,2}(t) \, x + q \, \nu_c \, t \right] = \sum \limits_{q = 0}^{2 \eta} b_q(t) \, \cos \left[ 2 \pi \, q \, f_{x,1,2}(t) \, x + q \, \nu_c \, t \right]
\end{equation}
where the temporal window is absorbed into $b_q(t)$. Inserting this expression of the nonlinear illumination intensity into the general form of the MP-SPIFI signal in Eq.~\eqref{spifi_signal}, we find:
\begin{eqnarray}
S(t) & = & \dirac{ \sum \limits_{q = 0}^{2 \eta} b_q(t) \, \cos \left[ 2 \pi \, q \, f_{x,1,2}(t) \, x + q \, \nu_c \, t \right] \, C(x) }_{\bf r}  \nonumber \\
	& = &  \sum \limits_{q = 0}^{2 \eta} b_q(t) \, \dirac{ \cos \left( 2 \pi \, q \, M \, \frac{n_2}{n_1} \, \Delta k \, \nu_r \, t \, x + q \, \nu_c \, t \right) \, C(x) }_{\bf r} = \sum \limits_{q = 0}^{2 \eta} S_q(t)
\label{spifi_sum}
\end{eqnarray}
Equation~\eqref{spifi_sum} illustrates how $2\eta$ images with varying spatial frequency support are collected in a single rotation of the modulation mask. Moreover, single-shot collection of all images is possible because the carrier frequency for each image scales with frequency support.  Consequently, images can be separated by simply filtering in the modulation frequency domain.

\section{Limitations of resolution enhancement}
The plane wave analysis from the previous section indicates that a resolution enhancement of 4$\times$ is possible with a second-order MP-SPIFI imaging modality.  Moreover, the analysis implies that the shape of the OTF for each SPIFI order is a top-hat function, meaning that there is no decay in the spatial frequency amplitude throughout the duration of the scan. In practice, the circular geometry of the pupil plane imposes limitations on the resolution enhancement attainable by apodizing the OTFs for each SPIFI order. This is intuitively understood by observing that the shape of each diffracted beam in the pupil plane is highly elliptical. Since each of the diffracted beams scans laterally across the  $x$ dimension of the pupil, the vertical spatial frequency support of the $j^{\mathrm{th}}$ diffracted order, $f_{y,j}$, is scan-time dependent, reducing the throughput energy of the diffracted beams with scan time as well as increasing the vertical spread in the line focus created in the object plane. Both of these effects cause a reduction in the peak intensity in the object plane, and hence impose strict limitations on the spatial frequency support.

\subsection{Energy transmission vs. scan time}
\begin{figure}[!ht]
\begin{center}
\resizebox{3.5in}{!}{\includegraphics{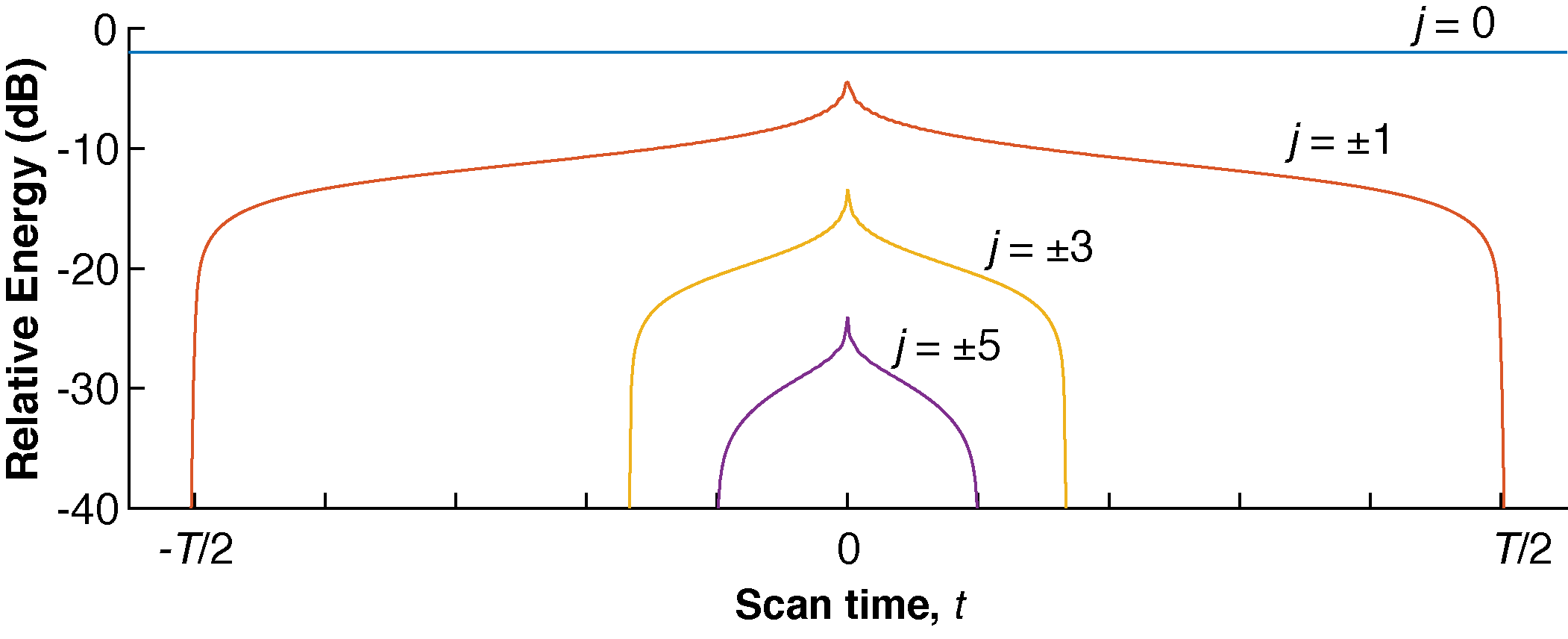}}
\caption{\label{sfig3} Simulated energy transmission of the diffracted orders through the objective lens. Relative energies are normalized by the total transmitted energy at scan time $t$=0, when all diffracted orders are overlaid in the lateral dimension.  }
\end{center}
\end{figure}

We examined the relative energy transmission as a function of scan time by numerically integrating amplitude of the diffracted fields described by the pupil function over the pupil aperture:
\begin{equation}
e_{j}(t) = \int \limits_{0}^{2 \pi} \mathrm{d}\phi \int \limits_{0}^{\alpha} \mathrm{d}\theta \, \sin \theta \, P_j(\theta,\phi,t)
\end{equation}
Here $P_j(\theta,\phi,t)$ is the time-dependent pupil function, describing the distribution of the electric field amplitude in the pupil plane.  The relative energy transmission curves are shown in Fig.~S\ref{sfig3}. Clearly the higher-order diffracted modes contribute less to the overall energy of the illumination intensity.

\subsection{Vignetting of diffracted illumination beams}
The circular pupil of the objective lens reduces the amplitude of high spatial frequencies in measured data. For example, the fourth-order MP-SPIFI image is formed by the second-order nonlinear response of the intensity pattern created by interference between the $j=\pm1$ beams:
\begin{equation}
S_{4+}(t) = \dirac{ I_{4+}({\bf r},t) \, C({\bf r},t) }_{\bf r} = \dirac{ \left[ v_{1,2}({\bf r},t) \, u_{-1,2}({\bf r},t) \right]^2 \,  C({\bf r},t)}_{\bf r}
\end{equation}
As these two beams scan off the optic axis of the imaging system, they undergo severe vignetting by the circular pupil of the objective lens (Fig.~S\ref{sfig1}d), causing a loss of transmitted energy and a decrease in the vertical spatial frequency support ($f_y$) that is responsible for a further decrease in the peak intensity in the object region.

We studied the effect of vignetting on lateral spatial frequency support by computing the illumination intensity in the object region with the angular spectrum representation \cite{Born:1999,Novotny:2012}. In order to compute $S(t)$, we must compute the illumination intensity in the object region. The vectorial electric field of each diffracted beam in the object region is:
\begin{equation}
{\bf v}_{j,2}({\bf r},t) = v_{x,j,2}({\bf r},t) \, {\bf x} + v_{y,j,2}({\bf r},t) \, {\bf y} + v_{z,j,2}({\bf r},t) \, {\bf z}
\end{equation}
and the illumination intensity becomes:
\begin{equation}
I_\mathrm{ill}({\bf r},t) = \sum \limits_{k = -N}^{N} \sum \limits_{j = -N}^{N} {\bf v}_{j,2}({\bf r},t) \cdot {\bf u}_{k,2}({\bf r},t) 
\end{equation}
The measured MP-SPIFI signal is then:
\begin{equation}
S(t) = \dirac{\left[ \sum \limits_{k = -N}^{N} \sum \limits_{j = -N}^{N} {\bf v}_{j,2}({\bf r},t) \cdot {\bf u}_{k,2}({\bf r},t) \right]^\eta \, C({\bf r},t)}_{\bf r}
\end{equation}

To obtain the illumination intensity at the focal plane, the Debye integral must be solved numerically for each diffracted order:
\begin{eqnarray}
\left( \begin{array}{c} v_{x,j,2}({\bf r},t) \\ v_{y,j,2}({\bf r},t) \\ v_{z,j,2}({\bf r},t) \end{array} \right) & = & \int \limits_{0}^{2 \pi} \mathrm{d} \phi \int \limits_{0}^{\alpha} \mathrm{d} \theta \, \sin \theta \, P_j(\theta,\phi,t) \, \left( 
	\begin{array}{c}
	\cos \phi \, \cos \theta \, \cos \left( \phi - \gamma \right) + \sin \phi \, \sin \left(\phi - \gamma \right) \\
	\sin \phi \, \cos \theta \, \cos \left( \phi - \gamma \right) - \cos \phi \, \sin \left(\phi - \gamma \right) \\
	\sin \theta \, \cos \left(\phi - \gamma \right)
	\end{array} \right)  \nonumber \\
	& & \times  \sqrt{\frac{n_1}{n_2} \cos \theta} \, \exp \left( \mathrm{i} \, k_2 \, z \, \cos \theta \right) \, \exp \left[\mathrm{i} \, k_2 \, \rho \, \sin \theta \, \cos \left(\phi - \varphi \right) \right]
\label{eq:debye}
\end{eqnarray}
In cartesian coordinates, the pupil function has the form of a two-dimensional Gaussian distribution:
\begin{equation}
P_j(x_p,y_p,t) = a_j \, \exp \left\{- \left[ \frac{ x_p - x_{0,j}(t)}{w_x}\right]^2 \right\} \exp \left\{- \left[ \frac{ y_p - y_{0,j}(t) }{w_y} \right]^2 \right\}
\label{eq:pupil_cart}
\end{equation}
where $(x_p,y_p)$ are coordinates in the pupil plane, $x_{0,j}(t)$ and $y_{0,j}(t)$ are the lateral and vertical shifts of the $j^{\mathrm{th}}$ diffracted order in the pupil plane, and $w_x$ and $w_y$ are the widths of the Gaussian distributions. To solve the integral equation in Eq.~\eqref{eq:debye}, the pupil function must be converted to spherical polar coordinates, $(F_o,\theta,\phi)$. Note that the radial coordinate is equivalent to the focal length of the objective lens because the pupil function is to be described on the surface of an aplanatic focal sphere \cite{Novotny:2012}. Applying the coordinate change, we find:
\begin{eqnarray}
x_p & = & F_0 \, \sin \theta \, \cos \, \phi \\
y_p & = & F_0 \, \sin \theta \, \sin \, \phi
\end{eqnarray}
and the pupil function in Eq.~\eqref{eq:pupil_cart} becomes:
\begin{equation}
P_j(\theta,\phi,t) = a_j \, \exp \left\{- \left[ \frac{F_o \, \sin \theta \, \cos \phi - x_{0,j}(t) }{w_x} \right]^2 \right\} \, \exp \left\{- \left[ \frac{F_o \, \sin \theta \, \sin \phi - y_{0,j}(t) }{w_y} \right]^2 \right\}
\end{equation}

The lateral shifts of the diffracted orders in the pupil plane are determined by the spatial frequency of the modulation mask as a function of scan time, the wavelength of the illumination light, and the focal length of the tube lens:
\begin{eqnarray}
x_{0,j}(t) & = & F_t \, j \, \frac{\lambda}{n_1} \, f_{x,1}(t) = F_t \, j \,\frac{\lambda}{n_1} \, \Delta k \, \nu_r \, t = j \, x_{0,1}(t) \\
y_{0,j} & = & F_t \, j \, \frac{\lambda}{n_1} \, f_{y,1} = F_t \, j \, \frac{\Delta k}{k_1} = j \, y_{0,1}
\end{eqnarray} 
The widths of the distributions are set by the focal lengths of the cylindrical and tube lenses, as well as the input beam size, $w_\mathrm{in}$, and the illumination wavelength:
\begin{eqnarray}
w_x & = & \frac{\lambda \, F_t}{ \pi \, w_\mathrm{in}}\\
w_y & = & \frac{F_t}{F_c} \, w_\mathrm{in}
\end{eqnarray}
where $F_c$ is the focal length of the cylindrical lens. 

\begin{figure}[!ht]
\begin{center}
\resizebox{!}{8.3in}{\includegraphics{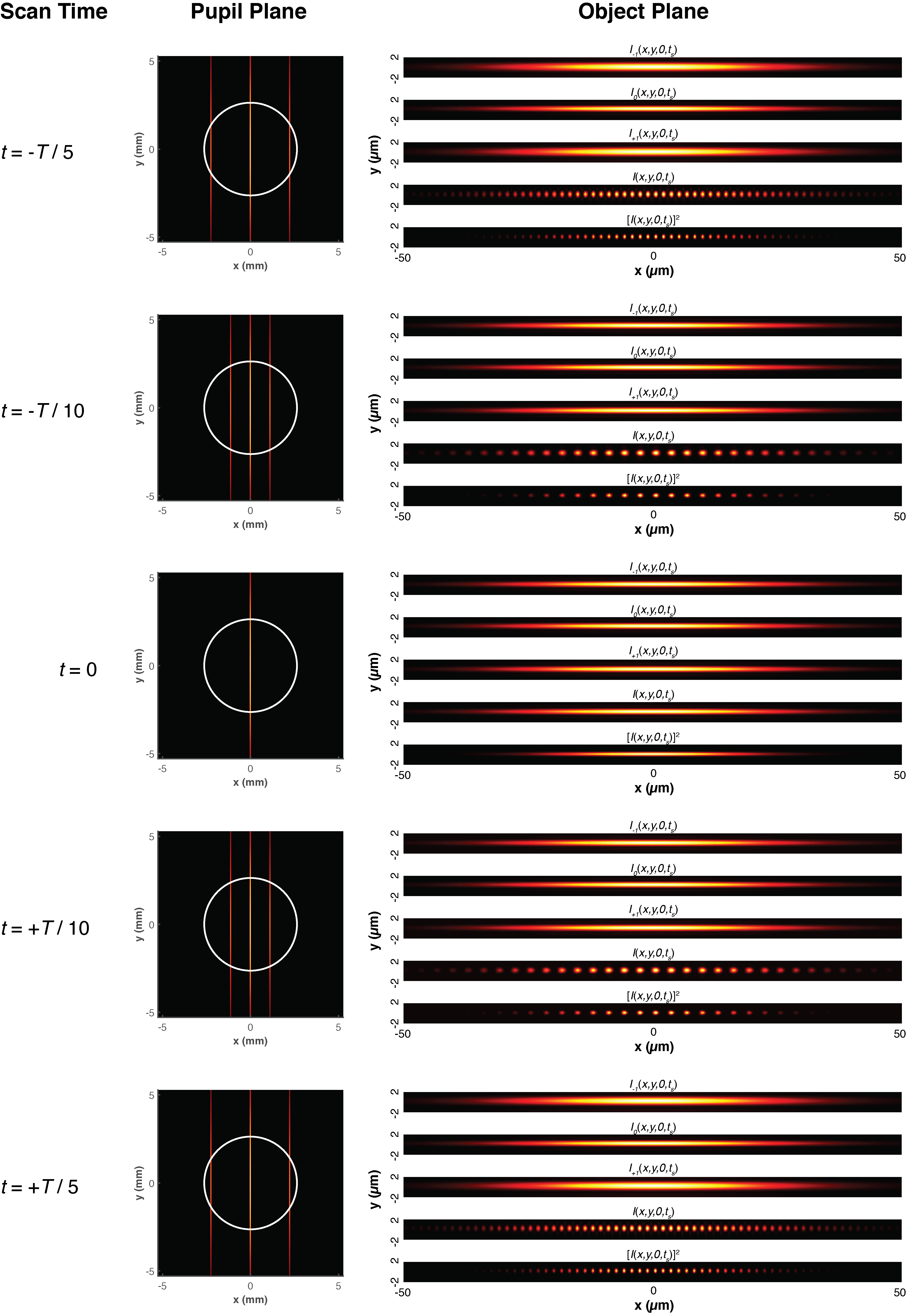}}
\caption{\label{fig:scan_summary} Intensities in the object and pupil planes, computed for several scan times with the angular spectrum representation.}
\end{center}
\end{figure}

We computed the illumination intensity in the transverse plane located in the focus of the illumination objective ($z=0$) for several scan times, considering only the $j = 0$ and $j=\pm1$ beams. From Fig.~S\ref{fig:scan_summary} it is clear that as the first-order diffracted beams scan away from the optic axis in the lateral ($x$) dimension, the vertical spatial frequency support decreases and causes a corresponding broadening of the illumination beams in the vertical dimension.

Finally, the vector-focusing formalism was used to compute the OTF of each MP-SPIFI order using the parameters of the microscope described in the Methods ($\lambda$=1064~nm; $\Delta k$=70/mm; 150~mm focal length tube lens; 50x/0.8~NA Zeiss air-immersion UIS series objective lens). To compute the MP-SPIFI signal, we assumed a point emitter located at the focal point of the objective lens, ${\bf r}_0 = (0,0,0)$. The point emitter was modeled by a Dirac-$\delta$ distribution, $\delta({\bf r} - {\bf r}_0)$. Since the contrast distribution we consider is a delta function, we need only compute the intensity at ${\bf r}_0$ to obtain the MP-SPIFI signal:
\begin{equation}
S(t) = \dirac{ \left[ I_\mathrm{ill}({\bf r},t) \right]^2 \, \delta ({\bf r} - {\bf r}_0)}_{\bf r} = \left[ I_\mathrm{ill}({\bf r}_0,t) \right]^2
\end{equation}
For the simulated TP-SPIFI data presented in Fig.~\ref{fig:tplsm_vs_tpspifi}, we considered a second-order nonlinearity ($\eta=2$), and included only first-order diffraction from the modulation mask, where $a_0 = 1/2$ and $a_1 = 1/\pi$ were set by the diffraction efficiency of a square grating.

\bibliography{refs3}

\end{document}